\documentclass{article} 
\usepackage{iclr2021_conference,times}

\usepackage{amsmath,amsfonts,bm}

\def\eqref#1{equation~\ref{#1}}

\def\1{\bm{1}}

\DeclareMathAlphabet{\mathsfit}{\encodingdefault}{\sfdefault}{m}{sl}
\SetMathAlphabet{\mathsfit}{bold}{\encodingdefault}{\sfdefault}{bx}{n}

\usepackage{hyperref}
\usepackage{url}
\usepackage{graphicx}
\usepackage[ruled,vlined,linesnumbered]{algorithm2e}

\usepackage{booktabs}       
\usepackage{amsfonts}       
\usepackage{nicefrac}       
\usepackage{microtype}      
\usepackage{mathtools}
\usepackage{ dsfont }
\usepackage{multirow}
\usepackage{textcomp}
\usepackage{cleveref}
\DeclarePairedDelimiter\abs{\lvert}{\rvert}%
\usepackage{amssymb}
\usepackage{wrapfig}
\usepackage{multicol}

\usepackage{enumitem}

\newcommand{\xhdr}[1]{\noindent{\bfseries #1.}}

\graphicspath{{figures/}}
\newcommand{\beginsupplement}{%
        \setcounter{table}{0}
        \setcounter{figure}{0}
        \setcounter{equation}{0}
        \renewcommand{\thetable}{S\arabic{table}}
        \renewcommand\thefigure{S\arabic{figure}}
        \renewcommand\theequation{S\arabic{equation}}
        \renewcommand{\theHtable}{Supplement.\thetable}
        \renewcommand{\theHfigure}{Supplement.\thefigure}
     }

\title{Neural representation and generation for RNA secondary structures}

\author{%
  Zichao Yan \\
  School of Computer Science\\
  McGill University, Mila\\
  \texttt{zichao.yan@mail.mcgill.ca} \\
  \And
  William L. Hamilton \\
  School of Computer Science\\
  McGill University, Mila\\
  \texttt{wlh@cs.mcgill.ca} \\
  \And
  Mathieu Blanchette \\
  School of Computer Science\\
  McGill University\\
  \texttt{blanchem@cs.mcgill.ca} \\
}

\iclrfinalcopy 
\begin{document}

\maketitle



\begin{abstract}
Our work is concerned with the generation and targeted design of RNA, a type of genetic macromolecule that can adopt complex structures which influence their cellular activities and functions. The design of large scale and complex biological structures {\color{black}spurs} dedicated graph-based deep generative modeling techniques, which represents a key but underappreciated aspect of computational drug discovery. In this work, we investigate the principles behind representing and generating different RNA structural modalities, and propose a flexible framework to jointly embed and generate these molecular structures along with their sequence in a meaningful latent space. Equipped with a deep understanding of RNA molecular structures, our most sophisticated encoding and decoding methods operate on the molecular graph as well as the junction tree hierarchy, integrating strong inductive bias about RNA structural regularity and folding mechanism such that high structural validity, stability and diversity of generated RNAs are achieved. Also, we seek to adequately organize the latent space of RNA molecular embeddings with regard to the interaction with proteins, and targeted optimization is used to navigate in this latent space to search for desired novel RNA molecules.
\end{abstract}

\section{Introduction}

There is an increasing interest in developing deep generative models for biochemical data, especially in the context of generating drug-like molecules.
Learning generative models of biochemical molecules can facilitate the development and discovery of novel treatments for various diseases, reducing the lead time for discovering promising new therapies and potentially translating in reduced costs for drug development~\citep{drug_discovery_pipeline}.
Indeed, the study of generative models for molecules has become a rich and active subfield within machine learning, with standard benchmarks~\citep{zinc}, a set of well-known baseline approaches~\citep{drug-vae,grammar_vae,cgvae,jtvae}, and high-profile cases of real-world impact~\footnote{e.g. \href{https://mila.quebec/en/ai-society/exascale-search-of-molecules}{LambdaZero} project for exascale search of drug-like molecules.}. 

Prior work in this space has focused primarily on the generation of small molecules (with less than 100 atoms), leaving the development of generative models for larger and more complicated biologics and biosimilar drugs (e.g., RNA and protein peptides) an open area for research. 
Developing generative models for larger biochemicals is critical in order to expand the frontiers of automated treatment design.
More generally, developing effective representation learning for such complex biochemicals will allow machine learning systems to integrate knowledge and interactions involving these biologically-rich structures. 

In this work, we take a first step towards the development of deep generative models for complex biomolecules, focusing on the representation and generation of RNA structures.
RNA plays a crucial role in protein transcription and various regulatory processes within cells which can be influenced by its structure~\citep{central_dogma,rna_biology}, and RNA-based therapies are an increasingly active area of research~\citep{mrna_vaccine,mrna_vaccine_2}, making it a natural focus for the development of deep generative models. 
The key challenge in generating RNA molecules---compared to the generation of small molecules---is that RNA involves a hierarchical, multi-scale structure, including a {\em primary sequential structure} based on the sequence of nucleic acids as well as more complex {\em secondary and tertiary structures} based on the way that the RNA strand folds onto itself. 
An effective generative model for RNA must be able to generate sequences that give rise to these more complex emergent structures. 

There have been prior works on optimizing or designing RNA sequences---using reinforcement learning or blackbox optimization---to generate particular RNA secondary structures~\citep{rnainverse_rl,rna-inverse-review}.
However, these prior works generally focus on optimizing sequences to conform to a specific secondary structure.
In contrast, our goal is to define a generative model, which can facilitate the sampling and generation of diverse RNA molecules with meaningful secondary structures, while also providing a novel avenue for targeted RNA design via search over a tractable latent space.

\xhdr{Key contributions}
We propose a series of benchmark tasks and deep generative models for the task of RNA generation, with the goal of facilitating future work on this important and challenging problem. 
We propose three interrelated benchmark tasks for RNA representation and generation:
\begin{enumerate}[leftmargin=*, itemsep=2pt, parsep=0pt]
    \item {\bf Unsupervised generation:} Generating stable, valid, and diverse RNAs that exhibit complex secondary structures.
    \item {\bf Semi-supervised learning:} Learning latent representations of RNA structure that correlate with known RNA functional properties.
    \item {\bf Targeted generation:} Generating RNAs that exhibit particular functional properties. 
\end{enumerate}
These three tasks build upon each other, with the first task only requiring the generation of stable and valid molecules, while the latter two tasks involve representing and generating RNAs that exhibit particular properties. 
In addition to proposing these novel benchmarks for the field, we introduce and evaluate three generative models for RNA.
All three models build upon variational autoencoders (VAEs)~\citep{vae} augmented with normalizing flows~\citep{vae_nf,iaf}, and they differ in how they represent the RNA structure. To help readers better understand RNA structures and properties, a self-contained explanation is provided in appendix~\ref{ap:rna-background}.

The simplest model (termed LSTMVAE) learns using a string-based representation of RNA structure. 
The second model (termed GraphVAE) leverages a graph-based representation and graph neural network (GNN) encoder approach~\citep{gnn-quantum-chem}. 
Finally, the most sophisticated model (termed HierVAE) introduces and leverages a novel hierarchical decomposition of the RNA structure. 
Extensive experiments on our newly proposed benchmarks highlight how the hierarchical approach allows more effective representation and generation of complex RNA structures, while also highlighting important challenges for future work in the area. 

\section{Task description}

Given a dataset of RNA molecules, i.e. sequences of nucleotides and corresponding secondary structures, our goals are to: (a) learn to generate structurally stable, diverse, and valid RNA molecules that reflect the distribution in this training dataset; (b) learn latent representations that reflect the functional properties {\color{black}of} RNA.
A key factor in both these representation and generation processes is that we seek to jointly represent and generate both the primary sequence structure as well as the secondary structure conformation. 
Together, these two goals lay the foundations for generating novel RNAs that satisfy certain functional properties. 
To meet these goals, we create two types of {\color{black}benchmark} datasets, each one focusing on one aspect of the above mentioned goals:

\xhdr{Unlabeled and variable-length RNA}
The first dataset contains unlabeled RNA with moderate and highly-variable length (32-512 nts), obtained from the human transcriptome~\citep{ensembl} and through which we focus on the generation aspect of structured RNA and evaluate the validity, stability and diversity of generated RNA molecules.
In particular, our goal with this dataset is to jointly generate RNA sequences and secondary structures that are biochemically feasible (i.e., valid), have low free energy (i.e., stable), and are distinct from the training data (i.e., diverse). 
We will give an extended assessment of the generation aspect under different circumstances, e.g., when constraining the generation procedures with explicit rules.

\xhdr{Labeled RNA}
The second dataset is pulled and processed from a previous study on {\color{black}\textit{in vitro}} RNA-protein interaction, which features labeled RNAs with shorter and uniform length (40 nts)~\citep{rnacompeteS}. With this dataset, our objective is slightly expanded (to include obj. a), so that the latent space is adequately organized and reflective of the interaction with proteins. Therefore, key assessment for the latent space includes AUROC for the classification of protein binding, which is crucial for the design of desired novel RNA molecules.

Essentially, this creates slight variations in the task formulation, with the first dataset suited to unsupervised learning of a generative model, while the second datasets involves additional supervision (e.g., for a semi-supervised model or targeted generation). Our specific modeling choices, to be introduced in section~\ref{sec:methods}, are invariant to different task formulations, and flexible enough to handle different representations of RNA secondary structures. We refer readers to appendix~\ref{ap:dataset_evaluation} for detailed explanation for the dataset and evaluation metrics on the generated molecules and latent embeddings.

\section{Methods} \label{sec:methods}

In this section, we introduce three different generative models for RNA. 
All three models are based upon the variational autoencoder (VAE) framework, involving three key components:

\begin{wrapfigure}{R}{0.48\textwidth}
\vspace{-1.5em}
\centerline{\includegraphics[width=0.48\textwidth]{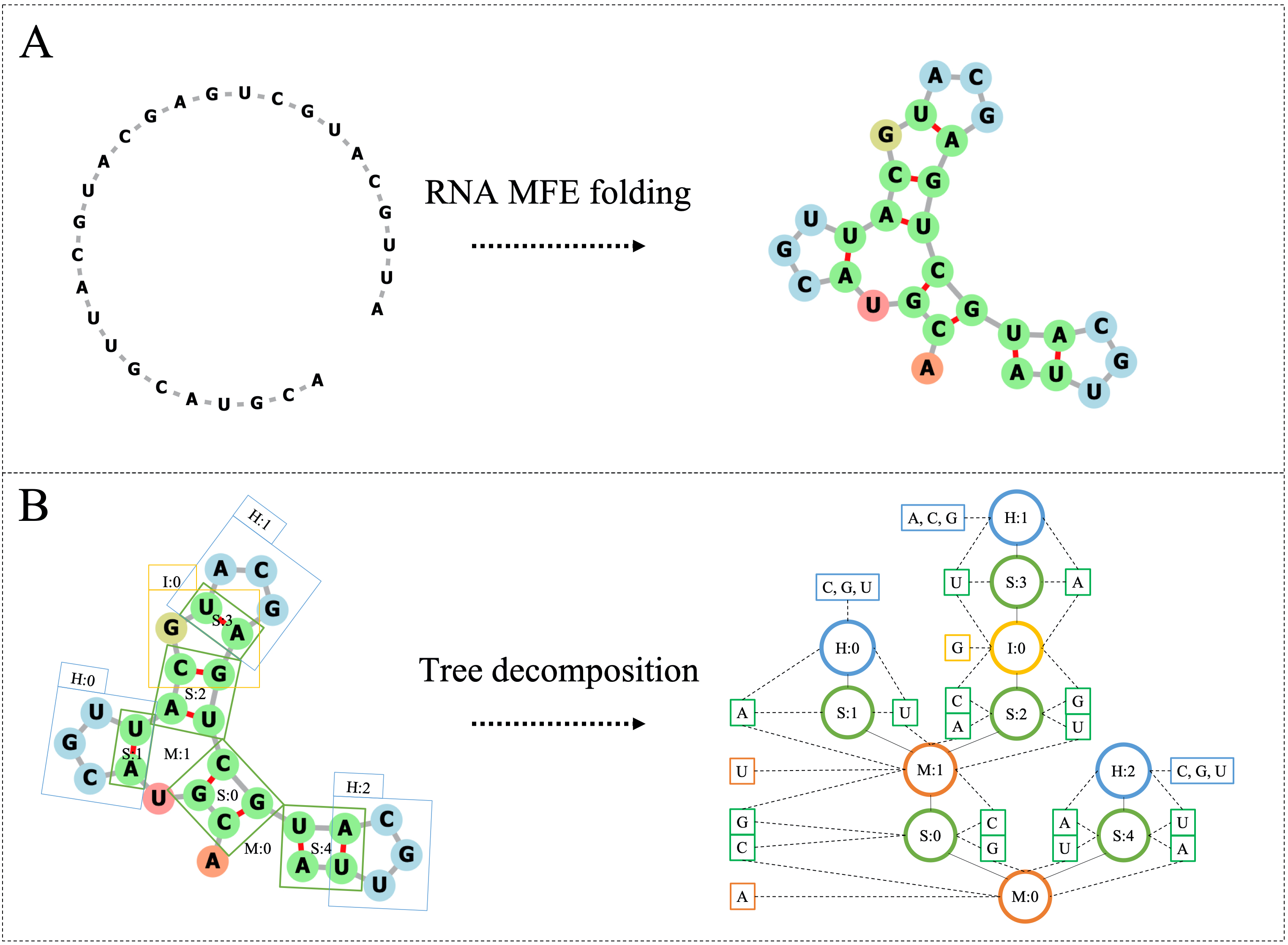}}
\caption{Hierarchical encoding. Panel (A) shows the minimum free energy structure (for illustration purpose only) which is represented as a graph. In panel (B), the original molecule is decomposed by grouping structurally adjacent nucleotides into {\color{black}subgraphs}, giving rise to a higher-order junction tree representation.}
\label{fig:encoding}
\vspace{-3em}
\end{wrapfigure}

\begin{enumerate}[leftmargin=*, itemsep=2pt, parsep=2pt, topsep=2pt]
    \item {\bf A probabilistic encoder network} $q_\phi(z | x)$, which generates a distribution over latent states given an input representation of an RNA. We experiment with three different types of input encodings for RNA sequence and secondary structures (see Figure \ref{fig:rna_rep}: a dot-bracket annotated string, a graph with adjacency matrix representing base-pairings, and a graph augmented with a hierarchical junction tree annotation for the secondary structure. 
    \item {\bf A probabilistic decoder network} $p_\theta(x | z)$, which defines a joint distribution over RNA sequences and secondary structures, conditioned on a latent input. As with the encoder network, we design architectures based on a linearized string decoding and a graph-based hierarchical junction-tree decoding approach. 
    \item {\bf A parameterized prior} $p_\psi(z)$, which defines a prior distribution over latent states and is learned based on a continuous normalizing flow (CNF)~\citep{cnf}. 
\end{enumerate}

For all the approaches we propose, the model is optimized via stochastic gradient descent to minimize the evidence lower bound (ELBO): $\mathcal{L} = -\mathbb{E}_{q_\phi(z | x)}[p_\theta(x | z)] + \beta\,\textrm{KL}(q_\phi(z | x) | p_\psi(z))$
where $\beta$ is a term to allow KL-annealing over the strength of the prior regularization. 

In the following sections, we explain our three different instantiations of the encoder (\cref{sec:encoding}), decoder (\cref{sec:decoding}), as well as our procedures to structurally constrain the decoding process using domain knowledge (\cref{sec:regularity-constraints}) and our procedures to avoid posterior collapse  (\cref{sec:posterior}). 

\subsection{Encoding RNA secondary structures}
\label{sec:encoding}

The input to the encoder is a structured RNA molecule, with its sequence given by an ordered array of nucleotides $x_1 \dots x_L$, with $x_i \in \{A,C,G,U\}$, where $L$ is the length of the sequence, and its secondary structure, either represented as (1) a dot-bracket string  $\mathcal{S}$ = $\dot{x}_1 \dots \dot{x}_L$ with $\dot{x}_i \in \{., (, )\}$; (2) or as a graph $\mathcal{G}$ with two types of edges — covalent bonds along the RNA backbone, and hydrogen bonds between the base-pairs~\footnote{We do not differentiate the number of hydrogen bonds, which can be different depending on the base-pairs. For example, G-C has three hydrogen bonds whereas A-U only contains two.}. We use $x_{uv}$ to denote edge features between nucleotides $u$ and $v$; (3) or as a hypergraph $\mathcal{T}$ — a depth-first ordered array of subgraphs $\hat{\mathcal{G}}_1 \dots \hat{\mathcal{G}}_D$ with $\mathcal{L}(\hat{\mathcal{G}}_i) \in \{S, H, I, M\}$ indicating the subgraph label, and $\mathcal{I}(\hat{\mathcal{G}}_i) = \{j|j\in \{1\dots L\}\}$ indicating the assignment of nucleotides to each subgraph.

\textbf{Encoding RNA secondary structure as sequence.} First, we obtain a joint encoding over the nucleotide and the dot-bracket annotation, using the joint sequence-structure vocabulary $\{A, C, G, U\} \times \{. , ( , )\}$.
Then, these one-hot encodings are processed by a stacked bidirectional LSTM~\citep{lstm}, followed by a multi-head self-attention module~\citep{transformer} to weigh different positions along the RNA backbone. A global max-pooling is used to aggregate the information into $h_{\mathcal{S}}$, and then we obtain mean $\mu_{\mathcal{S}}$ and log variance $\log\sigma_{\mathcal{S}}$ from $h_{\mathcal{S}}$ through linear transformations, and draw latent encoding $z_{\mathcal{S}}$ from $\mathcal{N}(\mu_{\mathcal{S}},\sigma_{\mathcal{S}})$ using the reparameterization trick~\citep{vae}.

\textbf{Learning graph representation of RNA secondary structure.} To encode the graph view $\mathcal{G}$ of an RNA secondary structure, we pass rounds of neural messages along the RNA structure, which falls into the framework of Message Passing Neural Network (MPNN) as originally discussed in~\citet{gnn-quantum-chem} and similarly motivated by~\citet{jtvae}.

For much longer RNAs, it is conceptually beneficial to pass more rounds of messages so that a nucleotide may receive information on its broader structural context. However, this may introduce undesired effects such as training instability and over-smoothing issues. Therefor , we combine our MPNN network with gating mechanism, which is collectively referred as the G-MPNN:
\noindent\begin{minipage}{.6\linewidth}
\begin{equation}
  \hat{v}_{uv}^{t-1} = \sigma(W^g_{local} [x_u\,||\,x_{uv}] + W^g_{msg} \sum_{w\in N(u)} v_{wu}^{t-1})
\end{equation}
\end{minipage}%
\begin{minipage}{.4\linewidth}
\begin{equation}
  v_{uv}^{t} = \mathrm{GRU}(\hat{v}_{uv}^{t-1}, v_{uv}^{t-1})\vphantom{\sum}
\end{equation}
\end{minipage}
where $[\dots||\dots]$ denotes concatenation, $\sigma$ denotes the activation function and GRU indicates the gated recurrent unit~\citep{gru}. Then, after $T$ iterations of message passing, the final nucleotide level embedding is given by: $h_u = \sigma(W^g_{emb}[x_u\,||\,\sum_{v\in N(u)}v_{vu}^{T}])$. Before pooling the nucleotide level embeddings into the graph level, we pass $h_1\dots h_L$ through a single bidirectional LSTM layer, obtaining $\hat{h}_1\dots \hat{h}_L$ at each step, and $h_g = \max(\{\hat{h}_i|i\in {1...L}\})$. The latent encoding $z_\mathcal{G}$ is similarly obtained from $h_\mathcal{G}$ using the reparameterization trick.

\textbf{Hierarchical encoding of the RNA hypergraph.} To encode the junction tree $\mathcal{T}$ of RNA, we employ a type of GRU specifically suited to tree-like structures, which has previously been applied in works such as GGNN~\citep{ggnn} and JTVAE~\citep{jtvae}. We refer to this tree encoding network as T-GRU, and the format of its input is shown in Figure~\ref{fig:encoding}.

One major distinction between our RNA junction tree and the one used for chemical compounds~\citep{jtvae} is that an RNA subgraph assumes more variable nucleotide composition such that it is impossible to enumerate based on the observed data. Therefore, we need to dynamically compute the features for each node in an RNA junction tree based on its contained nucleotides, in a hierarchical manner to leverage the nucleotide level embeddings learnt by G-MPNN.

Considering a subgraph $\hat{\mathcal{G}}_i$ in the junction tree $\mathcal{T}$, we initialize its node feature with: $x_{\hat{\mathcal{G}}_i}=[\mathcal{L}(\hat{\mathcal{G}}_i)\,||\max_{u \in \mathcal{I}(\hat{\mathcal{G}}_i)}h_u]$. Notably, $\max_{u \in \hat{\mathcal{G}}_i}h_u$ is a max-pooling over all nucleotides assigned to $\hat{\mathcal{G}}_i$, and nucleotide embedding $h_u$ comes from G-MPNN. To compute and pass neural messages between adjacent {\color{black}subgraphs} in the RNA junction tree $\mathcal{T}$, we use the T-GRU network in Eq.\ref{eq:tree-enc-msg}
\noindent\begin{minipage}{.55\linewidth}
\vspace{-1em}\begin{equation}
  v^{t}_{\hat{\mathcal{G}}_i, \hat{\mathcal{G}}_j} = \mathrm{T}\textup{-}\mathrm{GRU}(x_{\hat{\mathcal{G}}_i},
    \{v^{t-1}_{\hat{\mathcal{G}}_k, \hat{\mathcal{G}}_i}\,|\,\hat{\mathcal{G}}_k \in N(\hat{\mathcal{G}}_i)\})\label{eq:tree-enc-msg}
\end{equation}
\end{minipage}%
\begin{minipage}{.45\linewidth}
\begin{equation}
  h_{\hat{\mathcal{G}}_i} = \sigma(W^{t}_{emb} [x_{\hat{\mathcal{G}}_i}\,||\sum_{\hat{\mathcal{G}}\in N(\hat{\mathcal{G}}_i)}h_{\hat{\mathcal{G}}}])\label{eq:tree-node-emb}
\end{equation}
\end{minipage}
with details of T-GRU provided in the appendix~\ref{ap:t-gru}, and compute the embeddings for subgraphs with Eq.~\ref{eq:tree-node-emb}. 
Further, we obtain a depth-first traversal of the subgraph embeddings $h_{\hat{\mathcal{G}}_1} \dots h_{\hat{\mathcal{G}}_{D'}}$ which is also the order for hierarchical decoding to be discussed later. This ordered array of embeddings is processed by another bi-directional LSTM , and the final tree level representation $h_\mathcal{T}$ is again given by the max-pooling over the bi-LSTM outputs. Likewise, latent encoding $z_\mathcal{T}$ is obtained from $h_\mathcal{T}$.

\subsection{RNA molecular generation}
\label{sec:decoding}

\textbf{Decoding linearized sequence and structure.}
In this setting, the decoder simply autoregressively decodes a token at each step, from the joint sequence-structure vocabulary mentioned before in section~\ref{sec:encoding}, plus one additional symbol to signal the end of decoding. To simplify the design choice, we use a single-layered forward-directional LSTM, and its hidden state is initialized with the latent encoding $z$, which can be either $z_\mathcal{S}$, $z_\mathcal{G}$ or $z_\mathcal{T}$.

\begin{wrapfigure}{R}{0.5\textwidth}
\vspace{-2em}
\centerline{\includegraphics[width=0.5\textwidth]{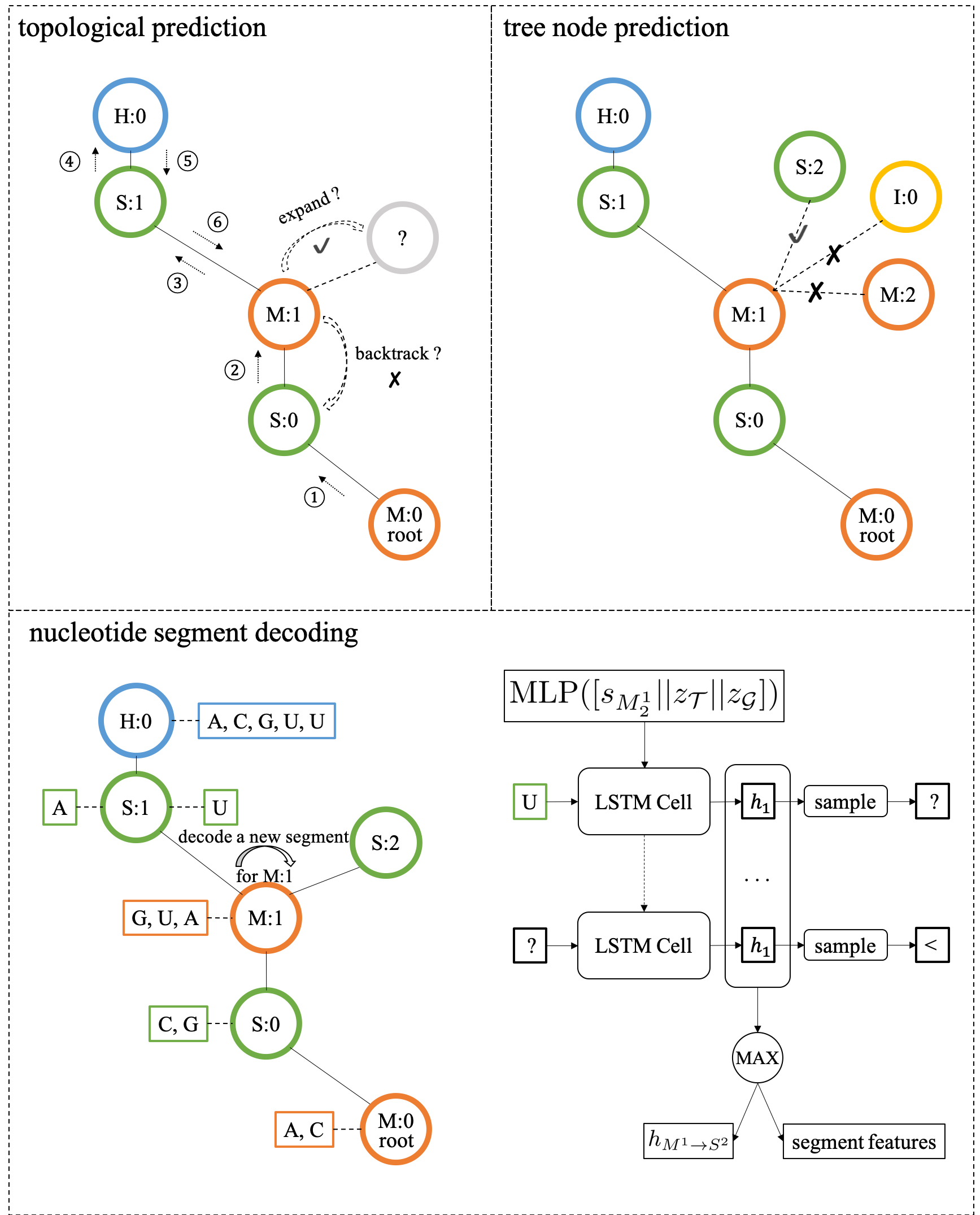}}
\caption{Hierarchical decoding of a structured RNA, involving three types of predictions, that are on the topological level, node level, and nucleotide level. These three types of prediction are interleaved into the procedures of decoding the junction tree structure of RNA and the nucleotide segments.}
\label{fig:decoding}
\vspace{-3em}
\end{wrapfigure}

\textbf{Hierarchically decoding hypergraph and nucleotide segments.} The input to this more sophisticated hierarchical decoder are latent encodings $z_\mathcal{G}$ which contains order and basic connectivity information of the nucleotides, and $z_\mathcal{T}$ which contains higher order information about the arrangements of nucleotide branches and their interactions. We give a concise description of the decoding procedures here, along with a detailed algorithm in appendix~\ref{ap:algo}. On a high level, we hierarchically decode the tree structure in a depth-first manner, and autoregressively generate a nucleotide segment for each visited tree branch. For these purposes, we interleave three types of prediction (Figure~\ref{fig:decoding}).

Denote the current tree node at decode step $t$ and at the $i$-th visit as $\hat{\mathcal{G}}^{t,i}$, whose features include (1) its node label $\mathcal{L}(\hat{\mathcal{G}}^{t,i})$ and, (2) a summary over the already existing $i-1$ nucleotide segments $\max\{h_u^{l,j}\,|\,u\in\hat{\mathcal{G}}^{t,i}\,\mathrm{and}\,l < t\,\mathrm{and}\,j < i\}$, with $l$ denoting the nucleotide is decoded at step $l$, and $j$ indicating the nucleotide belongs to the $j$-th branch (this feature is simply zeros when $i=1$). Then, its local feature $x_{\hat{\mathcal{G}}^{t,i}}$ is  defined as the concatenation of (1) and (2). 

We make use of a notion called \textit{node state}: $h_{\hat{\mathcal{G}}^{t,i}}$, which is obtained by: $h_{\hat{\mathcal{G}}^{t,i}}=\mathrm{T}\textup{-}\mathrm{GRU}(x_{\hat{\mathcal{G}}^{t,i}}, \{v_{\hat{\mathcal{G}}, \hat{\mathcal{G}}^{t,i}}\,|\,\hat{\mathcal{G}} \in N(\hat{\mathcal{G}}^{t,i})\})$. Note its similarity to Eq.~\ref{eq:tree-enc-msg}, and $h_{\hat{\mathcal{G}}^{t,i}}$ is used to make:
\begin{itemize}[leftmargin=*,itemsep=2pt, parsep=0pt]
    \item topological prediction in Figure~\ref{fig:decoding} (A), to determine if the decoder should expand to a new tree node or backtrack to its parent node, based on $\mathrm{MLP_{topo}}(h_{\hat{\mathcal{G}}^{t,i}})$;
    \item tree node prediction in Figure~\ref{fig:decoding} (B), on condition that a new tree node is needed due to a possible topological expansion. This procedure determines the label of the new tree node from the set of $\{S, H, I, M\}$, based on $\mathrm{MLP_{node}}(h_{\hat{\mathcal{G}}^{t,i}})$;
    \item nucleotide segment decoding in Figure~\ref{fig:decoding} (C), using a single-layered LSTM, whose initial hidden state is $\mathrm{MLP_{dec}}([h_{\hat{\mathcal{G}}^{t,i}}\,||\,z_\mathcal{T}\,||\,z_\mathcal{G}])$. The start token is the last nucleotide from the last segment. 
\end{itemize}

Our hierarchical decoder starts off by predicting the label of the root node using $z_\mathcal{T}$, followed by topological prediction on the root node and decoding the first nucleotide segment. The algorithm terminates upon revisiting the root node, topologically predicted to backtrack and finishing the last segment of the root node. The decoded junction tree naturally represents an RNA secondary structure that can be easily transformed to the dot-bracket annotation, 
and the RNA sequence is simply recovered by connecting nucleotide segments along the depth-first traversal of the tree nodes.

\subsection{Structurally constrained decoding}
\label{sec:regularity-constraints}

To better regulate the decoding process so that generated RNAs have valid secondary structures, a set of constraints can be added to the decoding procedures {\color{black}at the inference stage}. Essentially, a valid RNA secondary structure needs to observe the following rules: (1) base-pairing complementarity, which means only the canonical base-pairs and Wobble base-pairs are allowed, i.e. [A-U], [G-C] and [G-U]; (2) hairpin loop should have a minimum of three unpaired nucleotides, i.e. for any two paired bases at position $i$ and $j$, $\abs{i-j} > 3$; (3) each nucleotide can only be paired once, and overlapping pairs are disallowed. 

We will translate the above rules into specific and applicable constraints, depending on specific decoders. For the sake of space, we only give a broad remark and leave more details in the appendix. 

\textbf{Linearized decoding constraints.} Since the linearized decoder simply proceeds in an autoregressive fashion, constraints can be easily enforced in a way that at each step, a nucleotide with an appropriate structural annotation is sampled by making use of masks {\color{black} and re-normalizing the probabilities}. Likewise, a stop token can only sampled when all opening nucleotides have been closed. More details to follow in appendix~\ref{ap:linear-constr}.

\textbf{Hierarchical decoding constraints.} The specific set of constraints for hierarchical decoding is discussed in appendix~\ref{ap:hier-constr}. Overall, considering the different natures of the three associated types of prediction, each one should require a set of different strategies, which are once again applicable by adding proper masks before sampling. As shown in the algorithm in appendix~\ref{ap:algo}, the set of constraints are applied to line~\ref{alg:topo}, \ref{alg:node} and \ref{alg:segment} with marked asterisk.



\subsection{Avoiding posterior collapse}
\label{sec:posterior}

As discussed in a line of previous works, VAEs with strong autoregressive decoders are susceptible to posterior collapse, an issue where the decoder simply ignores the latent encoding of the encoder
~\citep{lagging_inference}. Therefore, to avoid posterior collapsing, we make use of a carefully chosen KL annealing schedule during training to help the encoder adapt its information content in the latent encoding and in coordination with the decoder. This schedule is detailed in section~\ref{sec:results}. We also learn a parameterized prior as suggested in~\citet{vae_lossy}, but using a CNF instead, following a similar implementation to~\citet{pointflow}, with details given in appendix~\ref{ap:cnf}.

{\color{black}Our KL annealing schedule is chosen based on empirical observations, as to our knowledge, there has yet to exist any principled methods of selecting such schedule. We have used diagnostic metrics such as mutual information~\citep{lagging_inference} and active units~\citep{active_units} along with a validation set to select a proper KL annealing schedule which is to be described later in section~\ref{sec:results}}

\section{Results}\label{sec:results}

We consider three modes of evaluation: (1) unsupervised RNA generation; (2) generation using semi-supervised VAE models and (3) targeted RNA design from an organized latent space. Results are presented below, and relevant hyperparameters can be found in Table~\ref{ap:hp-models}.

\begin{figure}[!tpb]
\vspace{-1em}
\centerline{\includegraphics[width=\textwidth]{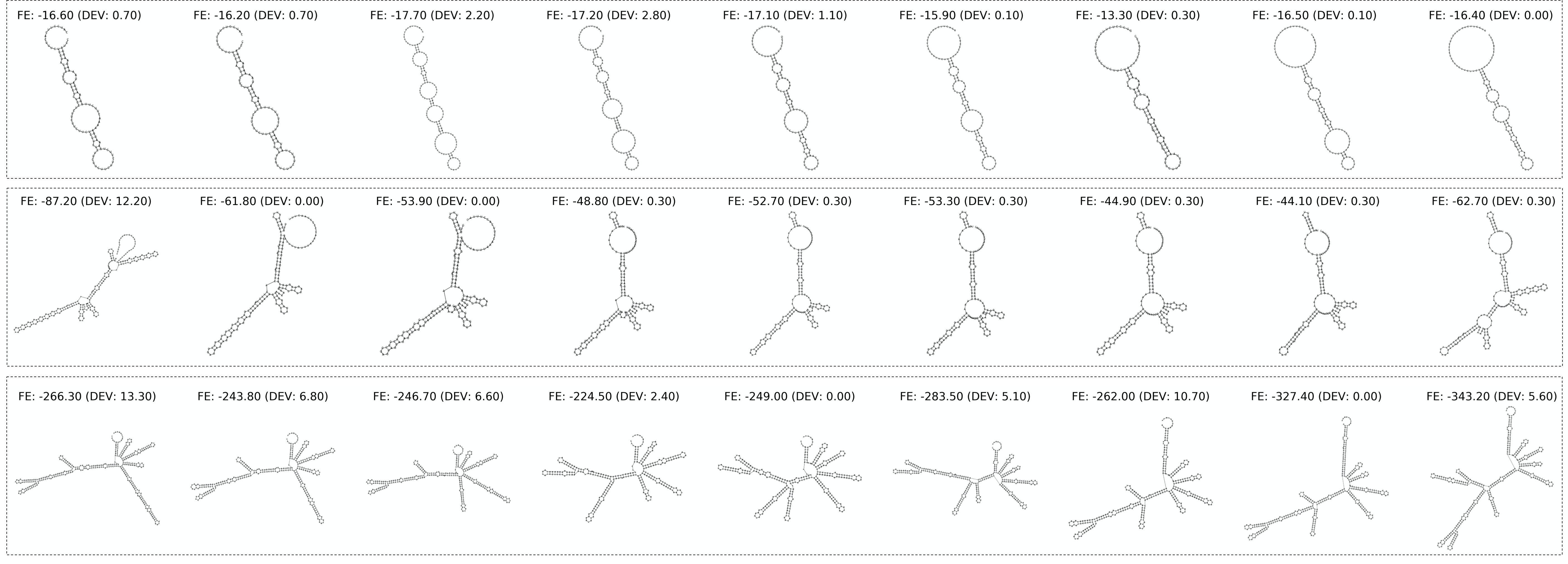}}
\caption{RNAs generated with structural constraints from HierVAE on a random axis in the latent space (step size: 1e-4), for short (top), medium-length (middle), and long (bottom) RNAs. The Free energy (FE) and its  deviation (DEV) from the MFE are given for each structure.}
\label{fig:prior-vis}
\vspace{-1em}
\end{figure}

\xhdr{Unsupervised RNA generation} Here, we evaluate generated RNAs from models trained on the unlabeled RNA dataset for 20 epochs using a KL annealing schedule including 5 epochs of warm-up, followed by gradually increasing the KL annealing term to 3e-3 (for LSTMVAE and GraphVAE), or 2e-3 (for HierVAE). 
The KL annealing schedule was chosen using a validation set of 1,280 RNAs.

Table~\ref{tb:post-prior-eval} compares the generation capability of different models, from the posterior as well as the prior distribution, and in scenarios such as applying structural constraints to the decoding process or not. It clearly shows that our most advanced model, HierVAE which employs a hierarchical view of the structure in its encoding/decoding aspects, achieves the best performance across different evaluation regimes, generating valid and stable RNAs even when the decoding processed is unconstrained. {\color{black} It is also observed that despite having structural constraints, the validity of our generated RNAs are always slightly below 100\%. This can be explained by the threshold hyperparameter which sets the maximum number of steps for topological prediction as well as the maximal length of each nucleotide segment, as shown in Algorithm 1 in appendix~\ref{ap:algo}.}

{\color{black}To further demonstrate the benefits of model training from structural constraints, we sample RNAs from the prior of an \textbf{untrained} HierVAE model. With structural constraints, the validity amounts to 66.34\% with an extremely high free energy deviation of 22.613. Without structural constraints, the validity translates to a mere 9.37\% and the model can only decode short single stranded RNAs as it lacks the knowledge of constructing more complex structures. This comparison illustrates that model training is essential for obtaining stable RNA folding.}

{\color{black}The junction tree hierarchy of RNAs developed in our work shares certain modelling similarities with the probabilistic context free grammar~\citep{cfg} used by covariance models (CM)~\citep{covariance_model}. Infernal~\citep{infernal} is one of the representative works based on CM, which is capable of sampling RNA secondary structures from a CM built around a consensus secondary structure for a conserved RNA family. However, due to the lack of homologous sequences in our dataset, Infernal is seriously limited and can only sample single stranded RNAs.}

Figure~\ref{fig:prior-vis} illustrate RNA structures generated using HierVAE from a randomly chosen short path through the latent space. Notably, latent encoding provided by HierVAE translates smoothly in the RNA structure domain: nearby points in the latent space result in highly similar, yet different, structures.  The generated structures are particularly stable for short and medium-size RNAs, and slightly less so for longer RNAs with highly complex structures. {\color{black} A side-by-side comparison between generated RNA secondary structures and MFE structures in Figure~\ref{fig:pairwise_struct} shows that generated structures can evolve smoothly in the latent space along with their corresponding MFE structures. We also visualize neighborhoods of a Cysteine-carrying transfer RNA and a 5S ribosomal RNA in figure~\ref{fig:trna} and~\ref{fig:ribosomal}.}

\begin{table}
\vspace{-1em}
  \caption{We evaluate RNAs sampled from the posterior distribution: $q(z|x)$, with a held-out test set of 20,000 RNAs. Each molecule is encoded and decoded 5 times. We also evaluate samples from the prior distribution: $\mathcal{N}(0,I)$ subject to the transformation of a latent CNF, where we sample 10,000 encodings and each encoding is decoded 10 times. Normed refers to length normalized FE DEV.}
  \label{tb:post-prior-eval}
  \centering
  \begin{tabular}{l@{\hspace{0.7\tabcolsep}}r@{\hspace{0.7\tabcolsep}}r@{\hspace{0.7\tabcolsep}}r@{\hspace{0.7\tabcolsep}}r@{\hspace{0.7\tabcolsep}}r@{\hspace{0.7\tabcolsep}}r@{\hspace{0.7\tabcolsep}}r@{\hspace{0.7\tabcolsep}}r}
    \toprule
    \multicolumn{1}{c}{} & 
    \multicolumn{4}{c}{Posterior Decoding} &
    \multicolumn{4}{c}{Prior Decoding} \\
    \cmidrule(r){2-5} \cmidrule(r){6-9}
    Model & Validity$\uparrow$ & FE DEV$\downarrow$ & Normed$\downarrow$ & Diversity$\uparrow$ & Validity$\uparrow$ & FE DEV$\downarrow$ & Normed$\downarrow$ & Diversity$\uparrow$\\
    \midrule
    \multicolumn{1}{c}{} & 
    \multicolumn{8}{c}{Constrained \& Stochastic} \\
    \midrule
    LSTMVAE & 99.47\% & 18.197 & 0.061 & 6.786 & 99.50\% & 18.536 & 0.062 & 6.789 \\
    GraphVAE & 99.47\% & 17.275 & 0.061 & 6.790 & 99.36\% & 18.534 & 0.065 & 6.791 \\
    HierVAE & 99.97\% & 8.678 & 0.035 & 6.787 & 99.86\% & 8.676 & 0.036 & 6.791 \\
    \midrule
    \multicolumn{1}{c}{} & 
    \multicolumn{8}{c}{Unconstrained \& Stochastic} \\
    \midrule
    LSTMVAE & 62.98\% & 8.700 & 0.048 & 6.791 & 62.58\% & 9.060 & 0.049 & 6.793 \\
    GraphVAE & 65.79\% & 9.508 & 0.051 & 6.792 & 63.45\% & 10.166 & 0.055 & 6.794 \\
    HierVAE & 94.51\% & 8.257 & 0.035 & 6.787 & 92.75\% & 7.897 & 0.037 & 6.791 \\
    \bottomrule
  \end{tabular}
  \vspace{-1em}
\end{table}

\begin{table}
  \vspace{-1em}
  \caption{Training semi-supervised HierVAE on labeled RNAcompete-S dataset. A test split is used to evaluate the accuracy of embedding classifiers and RNAs decoded from the posterior distribution under two settings: constrained and stochastic (C\& S), unconstrained and deterministic (NC\&D). {\color{black}RECON ACC refers to reconstruction accuracy which measures the percentage of RNA molecules decoded exactly as the input.}}
  \label{tb:su-vae-eval}
  \centering
  \begin{tabular}{lrrrrrrrr}
    \toprule
    \multicolumn{1}{c}{} & 
    \multicolumn{1}{c}{Test} &
    \multicolumn{3}{c}{Post C\&S} &
    \multicolumn{3}{c}{Post NC\&D}\\
    \cmidrule(r){3-5}\cmidrule(r){6-8}
    Dataset & AUROC & Valid & FE DEV & RECON ACC & Valid & FE DEV & RECON ACC\\
    \midrule
    HuR & 0.884 & 100\% & 0.426 & 55.85\% & 99.34\% & 0.269 & 68.73\% \\
    PTB & 0.907 & 100\% & 0.409 & 55.24\% & 92.07\% & 0.570 & 51.96\% \\
    QKI & 0.824 & 100\% & 0.439 & 55.80\% & 99.22\% & 0.296 & 66.83\% \\
    Vts1 & 0.900 & 100\% & 0.539 & 47.96\% & 98.90\% & 0.367 & 60.60\% \\
    RBMY & 0.878 & 100\% & 0.634 & 48.09\% & 98.64\% & 0.419 & 61.84\% \\
    SF2 & 0.900 & 100\% & 0.616 & 44.57\% & 98.88\% & 0.409 & 57.15\% \\
    SLBP & 0.792 & 100\% & 0.459 & 53.67\% & 98.60\% & 0.306 & 65.49\%  \\
    \bottomrule
  \end{tabular}
  \vspace{-1em}
\end{table}

\xhdr{Supervised RNA generation} We then evaluate our generative approaches in a semi-supervised setting using seven RBP binding data sets from RNAcompete-S. First, we compare the efficacy of different representational choices while excluding the generative components, i.e. we jointly train VAE encoders followed by simple MLP classifiers on top of the latent encodings for binary classification on RBP binding. 

Table~\ref{tb:full-classifier-eval} shows that incorporating RNA secondary structures is overall beneficial for the classification accuracy, except for RBMY where a model with access to RNA sequence alone (LSTM-SeqOnly) has the best performance. Notably, different choices for representing RNA secondary structures do not lead to large variation in performance, with the exception of HuR and SLBP, where graph based representations have an advantage over the linearized structural representation. {\color{black} On the other hand, sequence based models often have comparable performance, possibly due to the capability of inferring RNA secondary structures from short RNA sequences. It is also worth exploring other \textit{in-vitro} selection protocols such as HTR-SELEX which can select RNAs with higher binding affinities than RNAcompete-S that only involves a single selection step.}

Next, we train full generative models (encoder, decoder, latent CNF and MLP embedding classifier), and show the results in Table~\ref{tb:su-vae-eval}. Since our strategy for targeted RNA design makes use of seed molecules in the latent space, we mainly sample RNAs from the posterior distribution of these semi-supervised VAE models. Therefore, we select a KL annealing schedule that tends to retain more information in the latent encodings, i.e. setting maximum $\beta$ to 5e-4 and training 10 epochs.

Results are promising in that classification AUROC measured by the held-out test set is comparable to the fully supervised classification models in Table~\ref{tb:full-classifier-eval}, and much better compared to models only using fixed and pretrained VAE embeddings as shown in Table~\ref{tb:fixed-classifier-eval}. Also, RNA structures generated from the posterior distribution, even under the setting of unconstrained and deterministic decoding, have high success rates, very stable conformation and good reconstruction accuracy.

\begin{wraptable}{R}{0.4\textwidth}
\vspace{-1.5em}
\caption{Designing novel RNA with higher chances of RBP binding.}
\label{tb:rbp-opt}
\centering
\begin{tabular}{lrr}
\toprule
Dataset & Success & Improvement \\
\midrule
HuR & 96.88\% & 0.561$\pm$0.280 \\
PTB & 92.63\% & 0.561$\pm$0.320 \\
QKI & 90.82\% & 0.326$\pm$0.252 \\
Vts1 & 54.63\% & 0.197$\pm$0.388 \\
RBMY & 84.33\% & 0.457$\pm$0.395 \\
SF2 & 98.91\% & 0.655$\pm$0.239 \\
SLBP & 83.98\% & 0.309$\pm$0.297 \\
\bottomrule
\end{tabular}
\vspace{-1.5em}
\end{wraptable}
\xhdr{Targeted RNA design} We next studied the task of designing RNAs with high RBP binding affinity. Starting from the latent encodings of 10,000 randomly chosen RNA molecules that have negative labels in each RNAcompete-S test set, and use activation maximization to gradually alter the latent encodings so that the predicted binding probability from the embedding classifiers increases. These embedding classifiers have been trained jointly with the VAE models with accuracy reported earlier (Table~\ref{tb:su-vae-eval}). Then, we use separately trained full classifiers (also earlier shown in Table~\ref{tb:full-classifier-eval}) as proxy of oracles for evaluating the ``ground truth" probability of RBP binding.  Table~\ref{tb:rbp-opt}, report the success rate (fraction of RNAs whose ``ground truth" RBP binding probability was improved), along with the average improvement in binding probabilities. An example of a trajectory of optimized RNAs is shown in Fig. \ref{fig:targeted-gen-example}.

\section{Related work}
\label{ap:related_works}

Over the years, the field of computational drug discovery has witnessed the emergence of graph-centric approaches. One of the earliest method, proposed in~\citet{drug-vae}, is defined on the linearized format of molecular structures and represents a family of methods that rely on sequential models to represent and generate SMILES strings of chemical compounds. Later methods have sought to construct more chemical priors into the model, via (1) leveraging graph based representation and generation techniques, (2) enforcing direct chemical constraints to the decoding process, (3) considering a multi-scale view of the molecular structures, or (4) using reinforcement learning to integrate more training signal of the molecular structure and function. As a result, greater success has been achieved by models such as~\citet{grammar_vae,cgvae,jtvae,gcpn} at generating and searching valid and more useful chemical compounds.

Graph representation learning is at the heart of these more recent approaches, to help understand the rules governing the formation of these molecular structures, as well as the correspondence between structures and functions. 
\citet{gnn-mole-print} were among the first to apply GNN to learn molecular fingerprints, and the general neural message passing framework for molecules is proposed in~\citet{gnn-quantum-chem}, which demonstrate the power of MPNN across various molecular benchmarking tasks. These prior works on molecular MPNN, together with other GNN architectures developed in other areas, such as considering relational edges~\citep{rgcn} and attention~\citep{gat}, have laid the foundation for the success of these deep generative models.

Despite the fact that RNA molecules can adopt complex structures, dedicated graph representation learning techniques have been scarce, with some recent works beginning to leverage graph related learning techniques to predict RNA folding~\citep{transformer_rna,singh_dl_rna_folding} and to represent RNA molecular structures~\citep{yan_rna,carlos_rna_gnn}. Prior to our work, the design of RNA has mostly focused on the inverse design problem, which is to conditionally generate an RNA sequence whose MFE secondary structure corresponds to an input secondary structure. Therefore, the line of prior works have predominantly relied on sequential techniques, with some representative methods based on reinforcement learning~\citep{rnainverse_rl}, or more classically framed as a combinatorial optimization problem and solved with sampling based techniques~\citep{rna-inverse-review}. These prior works are mainly concerned with querying from an energy model with fixed thermodynamic parameters and fixed dynamics of RNA folding, which is in itself limited compared to learning based approaches~\citep{transformer_rna,singh_dl_rna_folding}, and are unable to model a joint distribution over RNA sequences and possible folds.

\section{Conclusion and future works}

In this work we propose the first graph-based deep generative approach for jointly embedding and generating RNA sequence and structure, along with a series of benchmarking tasks. Our presented work has demonstrated impressive performance at generating diverse, valid and stable RNA secondary structures with useful properties. 

{\color{black} For future works, there are several important directions to consider. First, it would be beneficial to obtain non-coding RNA families from the RFAM database
\citep{rfam} which would help our models learn more biologically-meaningful representation indicative of RNA homology and functions, in addition to the evolutionarily conserved RNA structural motifs that would enable the generation of more stable RNA secondary structures
. In that context, a detailed comparison to Infernal and other probabilistic context-free grammar models would be meaningful.

On the methodological aspect, in light of the recent advances in protein sequences pretraining across a large evolutionary-scale~\citep{facebook-protein,prottran}, our models for RNAs may similarly benefit by such a procedure with the data collected from RFAM. After the pretraining step, reinforcement learning can be used to finetune the generative component of our model with customizable rewards defined jointly on RNA structural validity, folding stability and functions such as binding to certain proteins. 

On the evaluation side, it would be of great interest to analyze our models for any potential RNA tertiary structural motifs and to compare them with those deposited in the CaRNAval~\citep{carnaval} or RNA 3D motifs database~\citep{rnamotifs}. Our models would also need modifications to allow non-canonical interactions and pseudoknots, which are common in RNA tertiary structures.

All in all, the representation, generation and design of structured RNA molecules represent a rich, promising, and challenging area for future research in computational biology and drug discovery, and an opportunity to develop fundamentally new machine learning approaches.
}

\newpage

\bibliography{articles.bib}
\bibliographystyle{iclr2021_conference}

\appendix
\beginsupplement
\onecolumn
\clearpage

\section{Acknowledgements}

We would like to thank all members of the Hamilton lab, Blanchette lab, and the four anonymous reviewers for their insightful suggestions. This work was funded by a Genome Quebec/Canada grant to MB and by the Institut de Valorisation des Donn\'ees (IAVDO) PhD excellence scholarship to ZY. WLH is supported by a Canada CIFAR AI Chair. We also thank Compute Canada for providing the computational resources.

\section{Background: RNA Structure and Key Properties}\label{ap:rna-background}
\begin{wrapfigure}{R}{0.4\textwidth}
\centering
\vspace{-1.2em}
\includegraphics[width=0.4\textwidth]{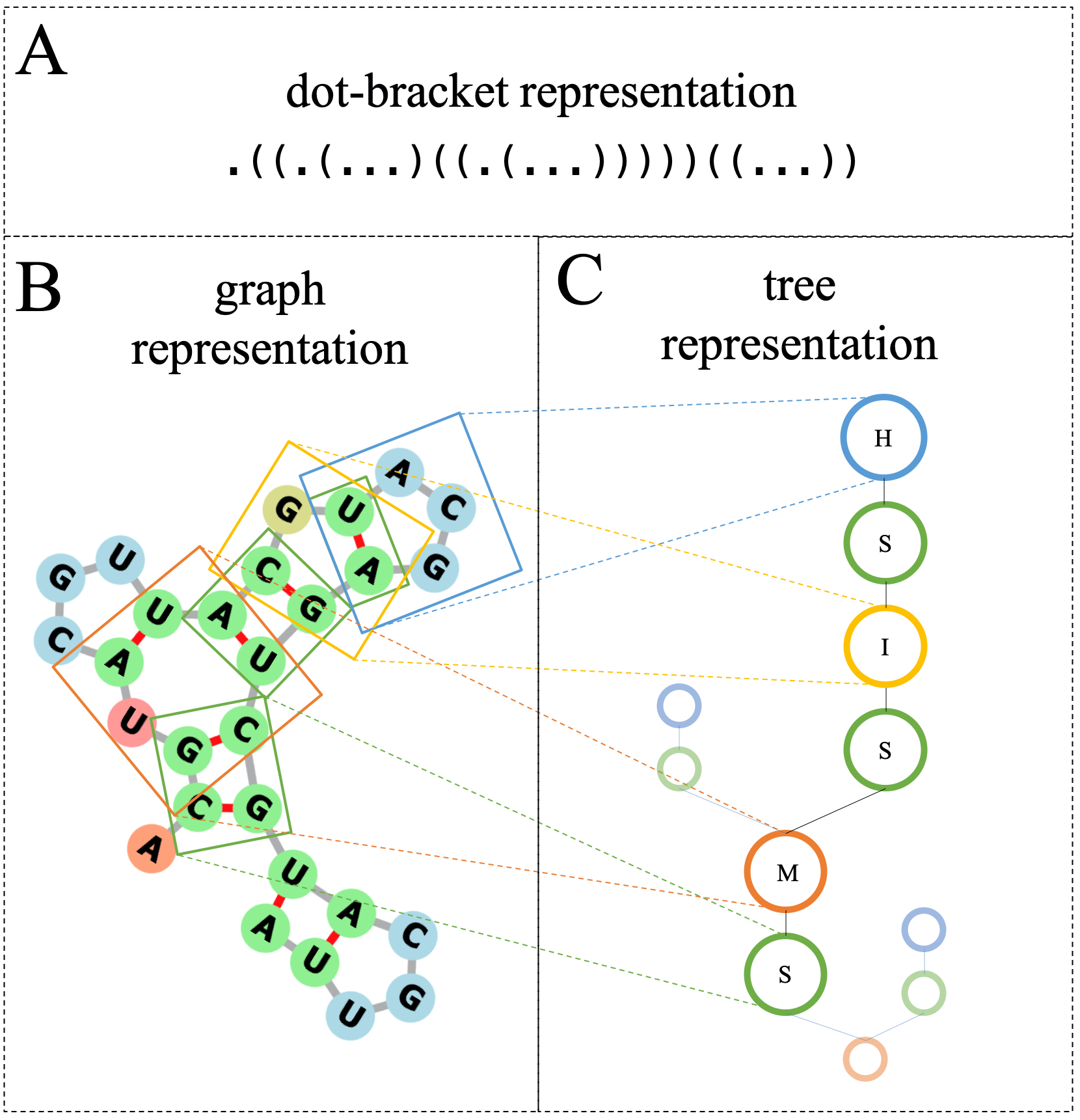}
\caption{\label{fig:rna_rep}A nested RNA secondary structure can be represented by: (A) dot-bracket annotation, where base-pairs corresponding to matching parentheses, or (B) a molecular planar graph with two types of edges, corresponding to consecutive nucleotides (backbone) and base-pairing interactions, or (C) a junction tree where node are labeled as stems (S), hairpins (H), internal loops (I), or multi-loops (M), and edges correspond to the connections between these elements.  All three forms are equivalent.}
\vspace{-2em}
\end{wrapfigure}

\textbf{The representation of an RNA molecule} starts from its {\em primary sequence structure}---i.e., a single chain of nucleotides (adenine (A), cytosine (C), guanine (G) and uracil (U)). 
RNA sequences are flexible and can fold onto themselves, enabling the formation of bonds between complementary nucleotides (Watson-Crick base-pairs [A-U, G-C], and Wobble base-pairs [G-U]), hence stabilizing the molecule \footnote{There exists other non-canonical base-pairs which are excluded from our current work.}. The set of pairs of interacting nucleotides in an RNA forms its so-called {\em RNA secondary structure}. In computational analyses of RNA, it is standard to assume that a secondary structure is {\em nested}: if $[i,j]$ and $[k,l]$ form base pairs with $i<k$, then either $l<j$ (nesting) or ${\color{black}k}>j$ (non-overlapping). This enables simple string or planar graph representations (Figure \ref{fig:rna_rep} a, b). 

The nested structure assumption means that secondary structures can be modelled by a probabilistic context free grammar~\citep{cfg}, or by the closely related junction tree structure (Figure \ref{fig:rna_rep} c)~\citep{roman-bayespairing2}, where each hypernode corresponds to a particular secondary substructure element: (1) {\em stem}: consecutive stacked base-pairs locally forming a double-stranded structure; (2) {\em hairpin loop }: unpaired regions closed by a base-pair; (3) {\em internal loop}: unpaired regions located between two stems; (4) {\em multiloop}: unpaired regions at the junction of at least three stems.
Edges link elements that are adjacent in the structure.

\textbf{Validity and stability of RNA folding.} The notion of free energy of RNA secondary structures can be used to characterize the stability of a particular conformation. Given an RNA sequence, there are combinatorially many valid RNA secondary structures which all need to obey a set of constraints (summarized in section~\ref{sec:regularity-constraints}). However, some structures are more stable than the others by having lower free energy. Therefore, these structures are more likely to exist (hence more useful) in reality due to the stochastic nature of RNA folding.
The free energy of an RNA secondary structure can be estimated by an energy-based model with thermodynamic parameters obtained from experiments~\citep{thermo_params}, wherein the minimum free energy (MFE) structure can be predicted, up to a reasonable approximation ~\citep{viennarna}.\footnote{Throughout this work, we use RNAfold~\citep{viennarna} to compute free energy as well as the MFE structure, due to its interpretability and acceptable accuracy for moderately sized RNAs.}

\newpage

\section{Dataset and Metrics} \label{ap:dataset_evaluation}

\begin{figure}[!h]
\centerline{\includegraphics[width=0.8\textwidth]{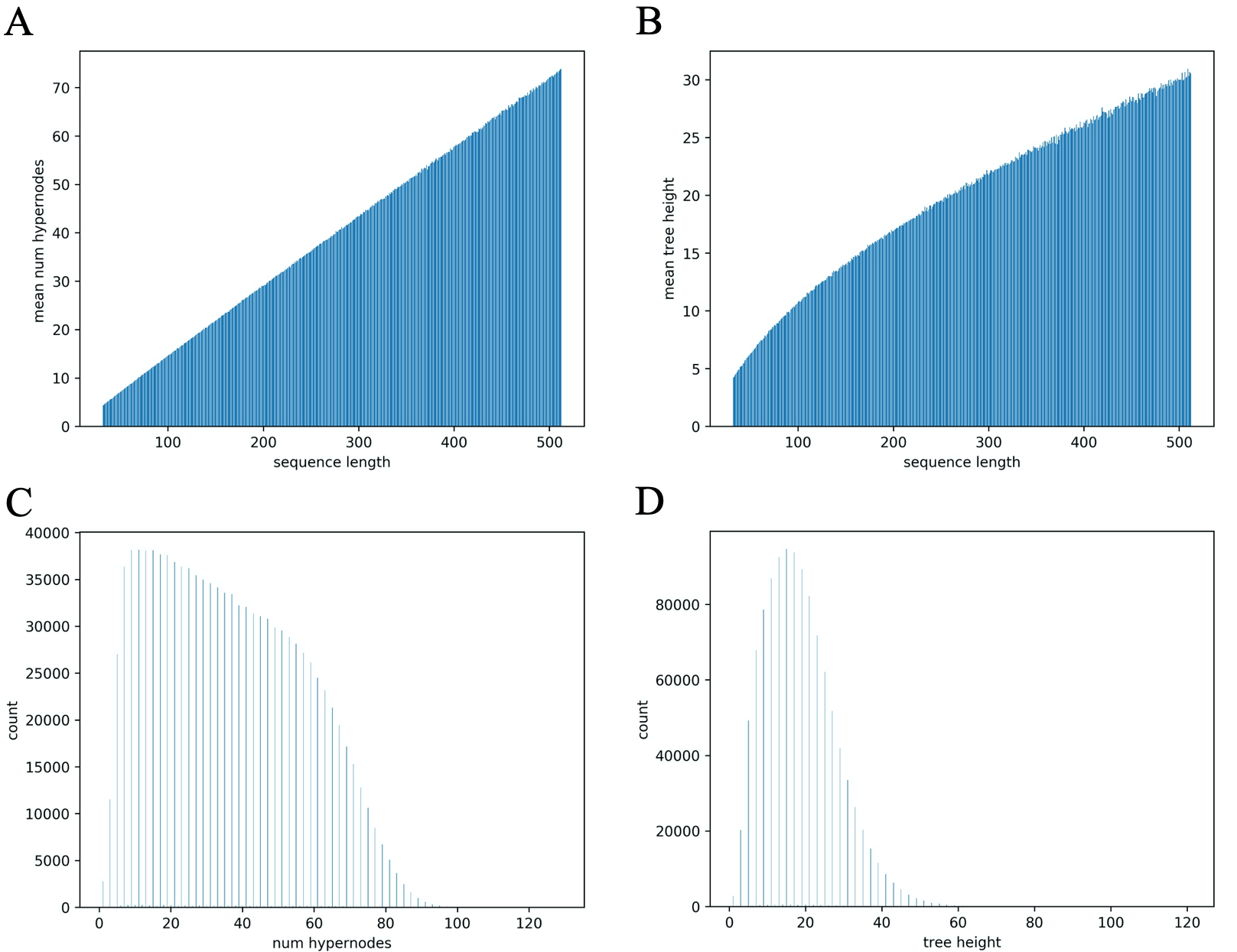}}
\caption{This figure contains information of the unlabeled RNA dataset. (A) The number of hypernodes appears to grow linearly with the length of RNA, and (B) the junction tree height also grows as the length increases but on a more moderate scale. (C) and (D) have shown bar-plots of the number of hypernodes and tree height, indicating that the junction tree of RNA can take on significant depth hence contributing to the diversity and complexity of RNA secondary structures represented in this dataset.}
\label{fig:dataset-info}
\end{figure}

The unlabeled dataset is obtained from the complete human transcriptome which is downloaded from the Ensembl database (\citet{ensembl}; version GRCh38). We slice the transcripts into snippets with length randomly drawn between 32 and 512 nts, and use RNAfold to obtain the MFE structures. We randomly split the dataset into a training set that contains 1,149,859 RNAs, and 20,000 held-out RNAs for evaluating decoding from the posterior distribution. More information on the structural diversity and complexity of this dataset is shown in Figure~\ref{fig:dataset-info}, which should present significant challenges for our algorithms.

The labeled dataset is pulled from a previous study on sequence and structural binding preference of RNA binding proteins (RBP), using an {\color{black}\textit{in vitro}} selection protocol called RNAcompete-S~\citep{rnacompeteS} which generates synthesized RNA sequences bound or unbound to a given RBP. RNAs in this experiment are of uniform length, i.e. 40 nts, and offer a rich abundance of RNA secondary structures compared to its predecessor protocols such as RNAcompete~\citep{rnacompete-protocol,rnacompete-dataset}. Since no benchmark has been ever established since its publication, we randomly sample 500,000 positive sequences bound to an RBP, and the same amount of negative sequences from the pool of unbound sequences, to curate a dataset for each of the seven RBPs investigated in the paper. Then, 80\% of all RNAs are randomly selected to the train split, and the rest goes to the test split.

Our evaluation scheme for the generated RNA secondary structures includes the following metrics:

\begin{itemize}[leftmargin=*,itemsep=2pt, parsep=0pt]
    \item \textbf{validity}: percentage of generated RNA secondary structures that conform to the structural constraints specified in section~\ref{sec:regularity-constraints}.
    \item \textbf{free energy deviation (FE DEV)}: difference of free energy between the generated RNA secondary structure and the MFE structure of the corresponding sequence, which quantifies the gap of both structures from an energy perspective. A lower FE DEV should indicate higher stability of generated RNAs. 
    \item \textbf{free energy deviation normalized by length (Normed FE DEV)}: FE DEV divided by the length of generated RNA, which distributes the contribution of total FE DEV to each base.
    \item \textbf{5-mer sequence diversity}: entropy of the normalized counts of 5-mer substrings, which directly measures the diversity of RNA sequences, and indirectly for RNA secondary structures when this metric is combined with FE DEV, since monolithic structures of diverse sequences would lead to high FE DEV.
\end{itemize}





\section{Tree encoding GRU}
\label{ap:t-gru}

Following Eq.\ref{eq:tree-enc-msg}, T-GRU computes a new message $v^t_{\hat{\mathcal{G}}_i,\hat{\mathcal{G}}_j}$ from $\hat{\mathcal{G}}_i$ and $\hat{\mathcal{G}}_j$, based on the features in $\hat{\mathcal{G}}_i$ denoted by $x_{\hat{\mathcal{G}}_i}$, as well as neural messages from neighboring {\color{black}subgraphs} to $\hat{\mathcal{G}}_i$, i.e. $\{v^{t-1}_{\hat{\mathcal{G}}_k,\hat{\mathcal{G}}_i}\,|\,\hat{\mathcal{G}}_k \in N(\hat{\mathcal{G}}_i)\}$. The internal structure of T-GRU is equivalent to the tree encoder employed in~\citet{jtvae}, which is essentially a neural analogue of the belief propagation algorithm on junction trees. Nevertheless, we write down the message passing formulas of T-GRU here:
\begin{align}
    s_{\hat{\mathcal{G}}_i,\hat{\mathcal{G}}_j} &= \sum_{\hat{\mathcal{G}}_k \in N(\hat{\mathcal{G}}_i)}v^{t-1}_{\hat{\mathcal{G}}_k,\hat{\mathcal{G}}_i} \\
    z_{\hat{\mathcal{G}}_i,\hat{\mathcal{G}}_j} &= \sigma(W^z [x_{\hat{\mathcal{G}}_i}\,||\,s_{\hat{\mathcal{G}}_i,\hat{\mathcal{G}}_j}] + b^z) \\
    r_{\hat{\mathcal{G}}_k,\hat{\mathcal{G}}_i} &= \sigma(W^r [x_{\hat{\mathcal{G}}_i}\,||\,v^{t-1}_{\hat{\mathcal{G}}_k,\hat{\mathcal{G}}_i}] + b^r) \\
    \hat{v}_{\hat{\mathcal{G}}_i,\hat{\mathcal{G}}_j} &= \textrm{Tanh}(W [x_{\hat{\mathcal{G}}_i}\,||\,\sum_{\hat{\mathcal{G}}_k \in N(\hat{\mathcal{G}}_i)} r_{\hat{\mathcal{G}}_k,\hat{\mathcal{G}}_i} \cdot v^{t-1}_{\hat{\mathcal{G}}_k,\hat{\mathcal{G}}_i}]) \\
    v^{t}_{\hat{\mathcal{G}}_i,\hat{\mathcal{G}}_j} &= (1 - z_{\hat{\mathcal{G}}_i,\hat{\mathcal{G}}_j}) \odot s_{\hat{\mathcal{G}}_i,\hat{\mathcal{G}}_j} + z_{\hat{\mathcal{G}}_i,\hat{\mathcal{G}}_j} \odot \hat{v}_{\hat{\mathcal{G}}_i,\hat{\mathcal{G}}_j}
\end{align}

\newpage

\section{Algorithm for hierarchically decoding structured RNA}
\label{ap:algo}

\begin{algorithm}[H]
\SetAlgoLined
\SetKwComment{tcc}{//\ }{}
\textbf{Given:} $z_\mathcal{T}$, $z_\mathcal{G}$, M\_TI, M\_SI\,\footnote{{\color{black}M\_TI refers to the threshold which set the maximum allowed number of topological prediction steps; M\_SI is another threshold to limit the length of each decoded nucleotide segment.}} \\
\textbf{Initialize:}  $stack \leftarrow [\,]$ \\
\SetKwProg{Fn}{function}{}{}
\Fn{decode($z_\mathcal{T}$, $z_\mathcal{G}$)}{
    \SetNoFillComment
    $root \leftarrow \mathrm{sample}(\mathrm{MLP}_{\mathrm{node}}(z_\mathcal{T}))$ \;
    $root$.add\_incoming\_message($z_\mathcal{T}$) \;
    $stack.\mathrm{push}((root, \,\mathbf{0}))$ \;
    $t \leftarrow 0$ \;
    \While{$t \leq \mathrm{M\_TI}\,\mathrm{and}\, stack.\mathrm{size}() \geq 1$}{
        $c\_node, last\_nuc \leftarrow stack$.get\_last\_item()\;
        $all\_msg \leftarrow \{msg\,|\,\forall msg \in c\_node\mathrm.\mathrm{get\_incoming\_message}() \}$ \;
        $local\_field \leftarrow [c\_node.\mathrm{label}()\,||\,c\_node.\mathrm{get\_segment\_features}()]$ \;
        $new\_msg \leftarrow \mathrm{T}\textup{-}\mathrm{GRU}(local\_field, all\_msg)$ \;
        \tcc{topological prediction}
        $is\_backtrack \leftarrow \mathrm{sample}(\mathrm{MLP}_{\mathrm{topo}}$($z_\mathcal{T}))^*$ \; \label{alg:topo}
        \tcc{nucleotide segment prediction}
        $new\_msg, last\_nuc, decoded\_segment, segment\_features \leftarrow decode\_segment(new\_msg, last\_nuc, z_\mathcal{T}$, $z_\mathcal{G}, \mathrm{M\_SI})^*$ \;
        \label{alg:segment}
        $c\_node.add\_decoded\_segment(decoded\_segment)$ \;
        $c\_node.add\_segment\_features(segment\_features)$ \;
        \eIf{$is\_backtrack = True$}
        {
            \tcc{backtrack to the parent node}
            $c\_node$.add\_incoming\_message($new\_msg$) \;
            $p\_node, \_ \leftarrow stack$.get\_penultimate\_item()\;
            $p\_node$.add\_neighbor($c\_node$) \;
            $stack$.update\_penultimate\_item(($p\_node, \,last\_nuc$))\;
            $stack$.pop$()$ \;
        }{
            \tcc{predict and expand to new tree node}
            $new\_node \leftarrow \mathrm{sample}(\mathrm{MLP}_{\mathrm{node}}(new\_msg))^*$ \;
            \label{alg:node}
            $new\_node$.add\_incoming\_message($new\_msg$) \;
            $new\_node$.add\_neighbor(($c\_node, \,last\_nuc$)) \;
            $stack$.push($new\_node$) \;
        }
        $t \leftarrow t + 1$ \;
    }
    return $root$ \;
}
\caption{DFS decode RNA secondary structure}
\end{algorithm}



\newpage

\section{Details for applying RNA structural constraints to linearized decoding procedures}
\label{ap:linear-constr}

When decoding from the joint vocabulary of sequence and dot-bracket structure ($\{A, C, G, U\} \times \{.,(,)\}$), whenever a nucleotide $\mathrm{nuc}_i$ with a left bracket is sampled at step $i$, we append them to a stack, i.e. $\{(\mathrm{nuc}_{i_0}, i_0)\dots (\mathrm{nuc}_i, i)\}$. Then, at decode step $j$,
\begin{itemize}[leftmargin=*,itemsep=2pt, parsep=0pt]
    \item if $\abs{i-j} \leq 3$, a proper mask will be added to the categorical logits of the vocabulary, to avoid sampling any nucleotides with right brackets, which means only an unpaired nucleotide or one that comes with a left bracket can be sampled;
    \item if $\abs{i-j} > 3$, a mask will be applied to make sure that only a nucleotide complementary to $\mathrm{nuc}_i$ can be sampled with the right bracket. Sampling nucleotides with other forms of structures are allowed.
\end{itemize}
As soon as a nucleotide with a closing right bracket is sampled, we pop out $(\mathrm{nuc}_i, i)$ from the stack. The special symbol for stop decoding can only be sampled when the stack has become empty.

\section{Details for applying RNA structural constraints to hierarchical decoding procedures}
\label{ap:hier-constr}

Additional constraints to be enforced during the hierarchical decoding process to ensure the validity of the decoded RNA secondary structure. Recall in section~\ref{sec:decoding} that three types of predictions are involved with the hierarchical decoding, therefore, each type is associated with its own set of rules. All set of rules can be observed by adding proper masks to the categorical logits before sampling, which are detailed below.

Constraints for making topological prediction, when the current node is
\begin{itemize}[leftmargin=*,itemsep=2pt, parsep=0pt]
    \item stem node, then the algorithm always expands to a new node upon its first visit, or backtracks to its parent node upon re-visit;
    \item hairpin node, then the algorithm always backtracks;
    \item internal loop, then the algorithm acts similarly as for stem node;
    \item multi-loop, then the algorithm always expands upon first visit and the next re-visit. Further re-visits to the same multi-loop node are not regulated. 
\end{itemize}

Constraints for predicting new tree node, when the current node is
\begin{itemize}[leftmargin=*,itemsep=2pt, parsep=0pt]
    \item stem node, then its child node when exists can be either a hairpin loop, an internal loop, or a multi-loop;
    \item hairpin node, internal loop or multi-loop, then its child node must be a stem node.
\end{itemize}

Constraints for decoding nucleotide segment. Due to the property of non-empty intersection between adjacent {\color{black}subgraphs}, the start token for decoding a segment at the current node, is always the last nucleotide decoded at the last node. Therefore, without explicitly mentioning, the algorithm needs to decode at least one new nucleotide at each segment. When the current node is
\begin{itemize}[leftmargin=*,itemsep=2pt, parsep=0pt]
    \item stem node, and if it is upon its first visit (i.e. decoding the first segment of a stem), then there is no for constraints. Otherwise, upon its re-visit, the algorithm needs to decode exactly the complementary bases and in the reverse order, according to the first decoded segment;
    \item hairpin node, then the decoder needs to decode at least four nucleotides before seeing the stop symbol, unless the hairpin is also the root node.
    \item internal loop node, and if it is upon its first, then constraint is not necessary. Otherwise, upon its revisit, the algorithm needs to decode at least one unpaired nucleotide on condition that the first decoded internal loop segment does not contain any unpaired nucleotides;
    \item multi-loop node, then there is no need for constraints.
\end{itemize}

\newpage

\section{Details for parameterizing prior distribution using normalizing flow}
\label{ap:cnf}

A normalizing flow involves a series of bijective transformation with tractable Jacobian log-determinant, to map an observed datapoint $x \sim p_\theta(x)$ from a complex distribution to a simpler one, such as the standard normal distribution.

Considering the simplified case where we have a single bijective function $f_\theta : \mathcal{Z} \rightarrow \mathcal{X}$ to map some simple latent variables $z$ to observed datapoint $x$, then, using the change of variable theorem, the likelihood of the observed datapoint can be evaluated as:
\begin{equation}
    p_\theta(x) = p_z(f^{-1}_\theta(x))|\mathrm{det}\frac{\partial f_\theta^{-1}(x)}{\partial x}|
\end{equation}
where $p_z(.)$ denotes some simple base distribution, e.g. $\mathcal{N}(0;I)$. Then, it becomes clear the efficiency of this scheme heavily relies on the efficiency of inverting the forward mapping $f_\theta$ as well as computing its Jacobian log-determinant.

In this project, we use a type of continuous normalizing flow (CNF) which simplifies the above mentioned computation~\citep{cnf}. Consider a time continuous dynamics $f_\psi(z(t),t)$ of some intermediate data representation $z(t)$, and again $z(t_0)\sim p_z(.)$, the transformation of variable, along with its inverse mapping, can be expressed as:
\begin{align}
    z \triangleq z(t_1) = z(t_0) + \int_{t_0}^{t_1} f_\psi(z(t),t)dt \\
    z(t_0) = z(t_1) + \int_{t_1}^{t_0} f_\psi(z(t),t)dt
\end{align}
and the change of probability density can be expressed as:
\begin{equation}
    \log p_\psi (z) = \log p_z(z(t_0)) - \int_{t_0}^{t_1}\mathrm{tr}(\frac{\partial f_\theta}{\partial z(t)})dt \label{eq:cnf_likelihood}
\end{equation}
Note that the invertibility issue is no longer a concern under some mild constraints~\citep{cnf}. Also, Eq.~\ref{eq:cnf_likelihood} only involves a more light-weight trace operation on the Jacobian rather than evaluating its log-determinant.

Therefore, we learn a parameterized prior using a CNF, and observe the decomposition of the KL term in the VAE objective:
\begin{equation}
    \mathrm{KL}(q_\phi(z|x)|p_\psi(z)) = -\mathbb{E}_{z\sim q_\phi(z|x)}[p_\psi(z)] - \mathbb{H}[q_\phi(z|x)] \label{eq:vae_kl}
\end{equation}
Therefore, during training our CNF parameterized with $\psi$ works on the transformation of complex latent encodings $z \sim q_\phi(z|x)$ to some simple $z(t_0) \sim \mathcal{N}(0;I)$, with an exact likelihood described by Eq.~\ref{eq:cnf_likelihood} and integrated into Eq.~\ref{eq:vae_kl} for the complete training objective. During inference, we simply sample $z_{t_0} \sim \mathcal{N}(0;I)$, and use our CNF to reversely transform it to $z \sim p_{\psi}(.)$ which should be closer to the approximate posterior.

Our specific parameterization of the CNF follows from \citet{pointflow} and \citet{ffjord}, interleaving two hidden \textit{concatsquash} layers of dimensionality 256 with Tanh non-linearity.

\newpage

\section{Information of the unlabeled RNA dataset}

\newpage
\section{Hyperparameters}

\begin{table*}[!h]
  \caption{Hyperparameters for training VAE and full classifier models. Note that hidden units refer to the dimensionality of encoders and decoders from LSTMVAE, GraphVAE as well as HierVAE models. Dropout is applied to the embedding MLP classifier in case of training semi-supervised VAEs, which contains one hidden layer.}
  \label{ap:hp-models}
  \centering
  \begin{tabular}{ll}
    \toprule
    \multicolumn{2}{c}{for VAE models} \\
    \midrule
    latent dimensionality & 128 \\
    hidden units & 512 \\
    G-MPNN iterations & 5 \\
    T-GRU iterations & 10 \\
    learning rate & 1e-3 \\
    batch size & 32 \\
    optimizer & AMSGrad~\citep{amsgrad}\\
    dropout ratio & 0.2 \\
    M\_TI & 300 \\
    S\_TI (hierarchical decoder) & 100 \\
    S\_TI (linearized decoder) & 1000 \\
    \midrule
    \multicolumn{2}{c}{\shortstack{for full classifier models\\(overriding some above hyperparameters)}} \\
    \midrule
    learning rate & 2e-4 \\
    epochs & 200 \\
    early stopping epochs & 5 \\
    \bottomrule
  \end{tabular}
\end{table*}

\section{RNAcompete-S classifiers on pretrained and fixed VAE embeddings}

\begin{table}[!h]
  \caption{Performance of simple MLP classifiers on top of fixed latent embeddings from VAE models, which have been pretrained on the unlabeled RNA dataset as originally shown in Table~\ref{tb:post-prior-eval}.}
  \label{tb:fixed-classifier-eval}
  \centering
  \begin{tabular}{lccc}
    \toprule
    RBP & LSTMVAE & GraphVAE & HierVAE \\
    \midrule
    HuR & 0.867 & 0.858 & 0.860\\
    PTB & 0.886 & 0.878 & 0.883\\
    QKI & 0.748 & 0.756 & 0.746\\
    Vts1 & 0.775 & 0.758 & 0.774\\
    RBMY & 0.734 & 0.725 & 0.731\\
    SF2 & 0.867 & 0.862 & 0.866\\
    SLBP & 0.749 & 0.737 & 0.747\\
    \bottomrule
  \end{tabular}
\end{table}

\newpage
\section{End-to-end RNAcompete-S classifiers}
\begin{table}[!h]
  \caption{We use the same encoding architectures as in the generative models, and report their AUROC averaged across 6 runs, for each RNAcompete-S RBP dataset.}
  \label{tb:full-classifier-eval}
  \centering
  \begin{tabular}{lcccc}
    \toprule
    RBP & LSTM-SeqOnly & LSTM & Graph & Hierarchical \\
    \midrule
    HuR & 0.880 $\pm$ 0.000 & 0.880 $\pm$ 0.000 & 0.880 $\pm$ 0.000 & \textbf{0.888 $\pm$ 0.002}\\
    PTB & 0.900 $\pm$ 0.000 & 0.910 $\pm$ 0.000 & 0.910 $\pm$ 0.000 & 0.910 $\pm$ 0.000 \\
    QKI & 0.820 $\pm$ 0.000 & 0.830 $\pm$ 0.000 & 0.825 $\pm$ 0.002 & 0.830 $\pm$ 0.000 \\
    Vts1 & 0.900 $\pm$ 0.000 & 0.908 $\pm$ 0.002 & 0.637 $\pm$ 0.079 & 0.910 $\pm$ 0.000\\
    RBMY & \textbf{0.905 $\pm$ 0.002} & 0.880 $\pm$ 0.003 & 0.802 $\pm$ 0.055 & 0.870 $\pm$ 0.002 \\
    SF2 & 0.890 $\pm$ 0.000 & 0.900 $\pm$ 0.000 & 0.900 $\pm$ 0.000 & 0.900 $\pm$ 0.000 \\
    SLBP & 0.777 $\pm$ 0.002 & 0.790 $\pm$ 0.000 & \textbf{0.797 $\pm$ 0.002} & \textbf{0.797 $\pm$ 0.002} \\
    \bottomrule
  \end{tabular}
\end{table}

\section{Alternative HierVAE training on RNAcompete-S}

\begin{table}[!h]
  \caption{Training HierVAE on supervised RNAcompete-S dataset. All models are trained with 20 epochs, including 5 epochs for warm-up, 6 epochs to linearly raise beta from 0 to 3e-3, and 9 remaining epochs with beta fixed at 3e-3. The test set measures AUROC and posterior decoding on the final model.}
  \label{tb:su-vae-alt-eval}
  \centering
  \begin{tabular}{lrrrrrrrr}
    \toprule
    \multicolumn{1}{c}{} & 
    \multicolumn{1}{c}{Test} &
    \multicolumn{3}{c}{Post R\&S} &
    \multicolumn{3}{c}{Post NR\&D}\\
    \cmidrule(r){3-5}\cmidrule(r){6-8}
    Dataset & AUROC & Valid & FE DEV & RECON ACC & Valid & FE DEV & RECON ACC\\
    \midrule
    HuR & 0.871 & 100\% & 0.951 & 18.97\% & 99.34\% & 0.702 & 31.52\% \\
    PTB & 0.899 & 100\% & 0.826 & 21.17\% & 98.64\% & 0.674 & 31.28\% \\
    QKI & 0.822 & 100\% & 0.867 & 17.82\% & 99.40\% & 0.627 & 30.62\% \\
    Vts1 & 0.874 & 100\% & 1.056 & 13.71\% & 99.39\% & 0.770 & 24.97\% \\
    RBMY & 0.872 & 100\% & 0.963 & 11.86\% & 98.68\% & 0.690 & 22.91\% \\
    SF2 & 0.874 & 100\% & 0.921 & 14.44\% & 99.32\% & 0.668 & 25.99\% \\
    SLBP & 0.764 & 100\% & 1.033 & 14.84\% & 99.44\% & 0.743 & 26.93\%  \\
    \bottomrule
  \end{tabular}
\end{table}

\newpage
\section{{\color{black}Comparison of generated RNA secondary structures to MFE structures}}
\begin{figure}[!h]
\centerline{\includegraphics[width=0.95\textwidth]{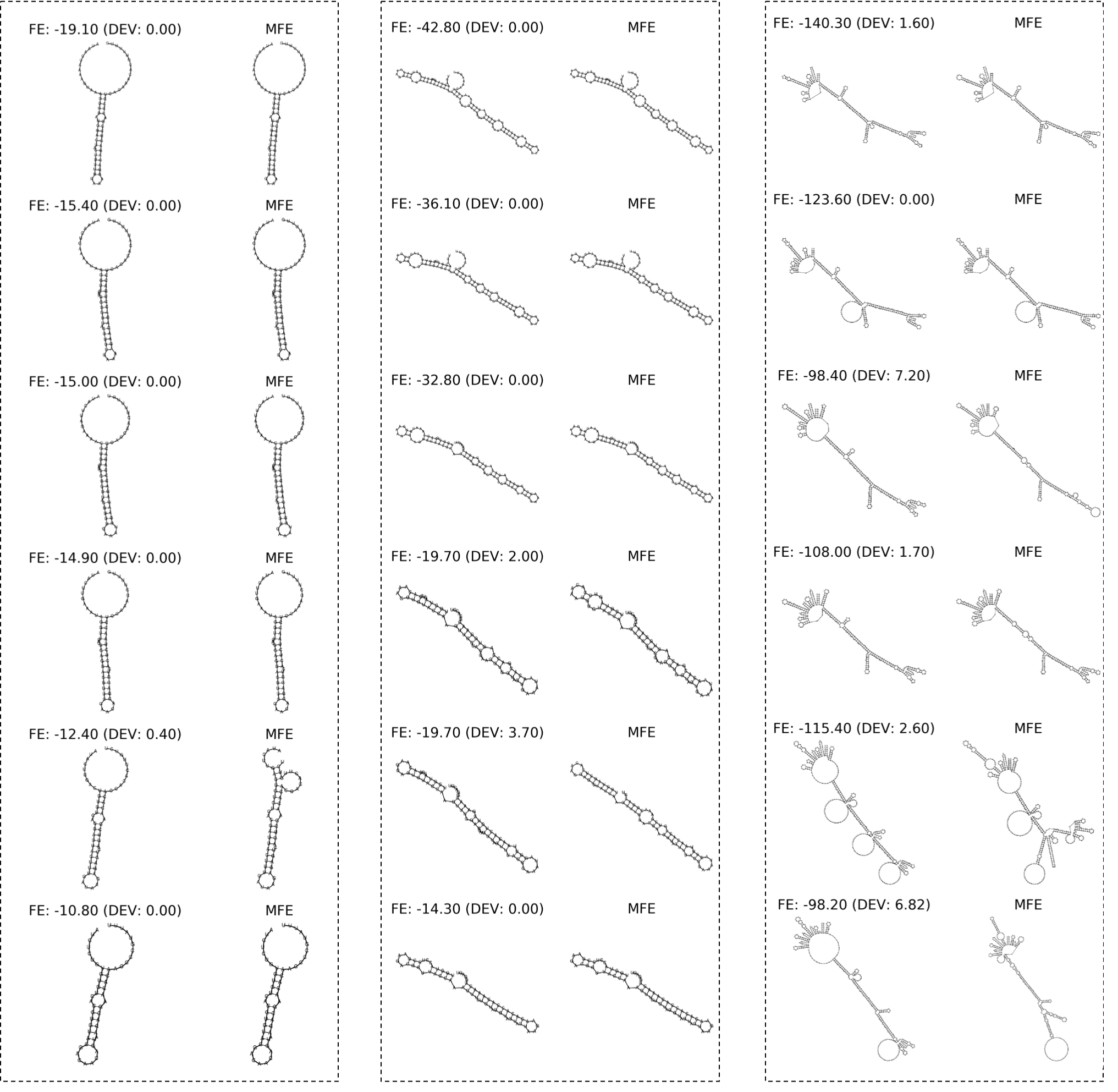}}
\caption{A comparison of generated RNAs (left) to their corresponding MFE structures (right). RNAs are generated with structural constraints from HierVAE on three random axes. The ground truth MFE structures are predicted by RNAfold, and the generated RNAs are shown to evolve smoothly in the latent space along with their corresponding MFE structures which have also shown relatively smooth transitions.}
\label{fig:pairwise_struct}
\end{figure}

\newpage
\section{{\color{black}Neighborhood visualization of a Cysteine-carrying transfer-RNA}}
\begin{figure}[!h]
\centerline{\includegraphics[width=0.95\textwidth]{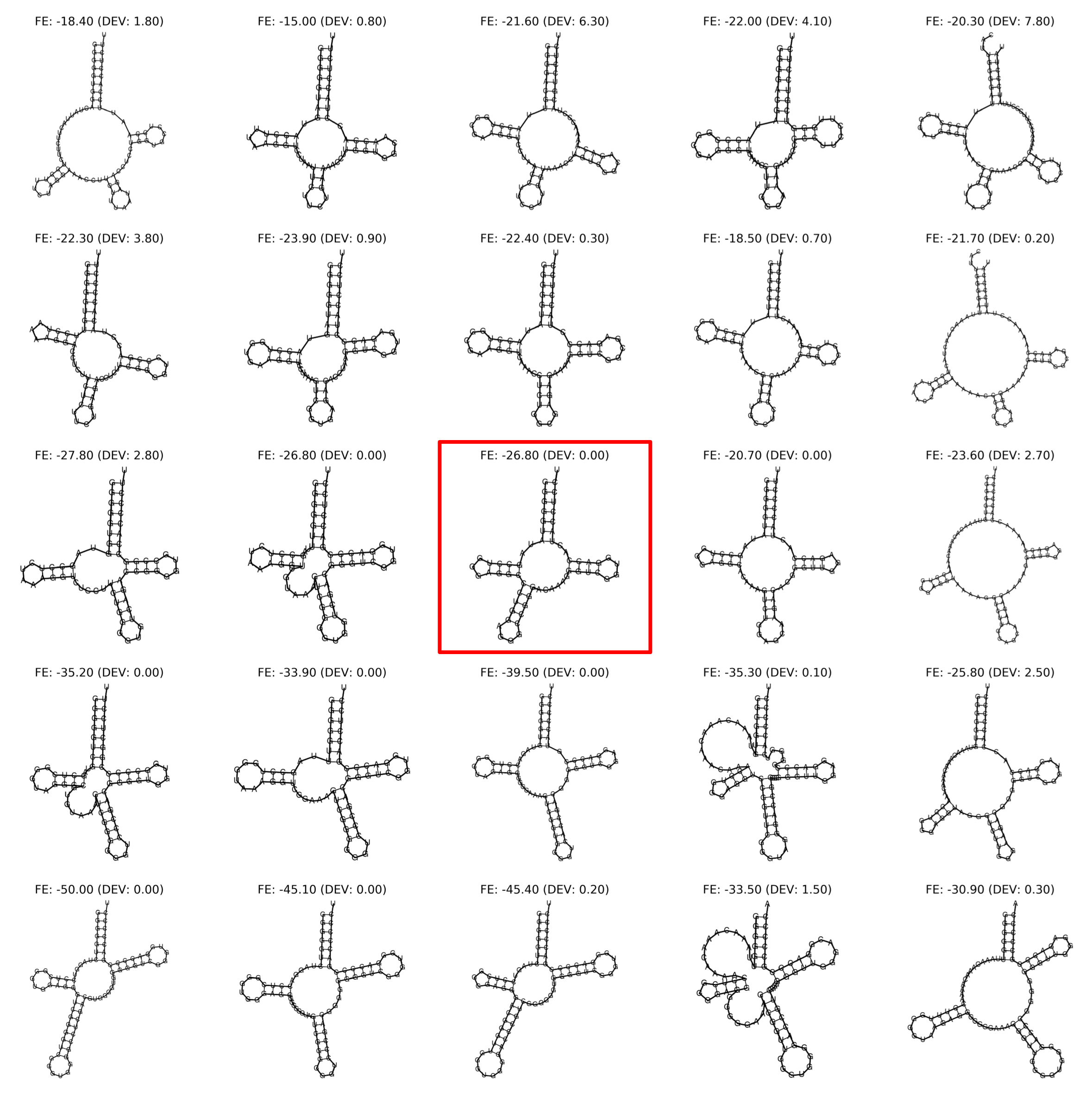}}
\caption{Neighborhood visualization of tRNA-Cys\protect\footnotemark which is marked by the red bounding box in the center and the walk in the latent space takes place on two random orthogonal axes. Note that actual secondary structure of tRNA-Cys plotted in the figure is different compared to the one deposited online due to the prediction of RNAfold.}
\label{fig:trna}
\end{figure}
\footnotetext{\href{https://rnacentral.org/rna/URS00001F47B5/9606}{https://rnacentral.org/rna/URS00001F47B5/9606}}

\newpage
\section{{\color{black}Neighborhood visualization of a 5S Ribosomal RNA}}
\begin{figure}[!h]
\centerline{\includegraphics[width=0.95\textwidth]{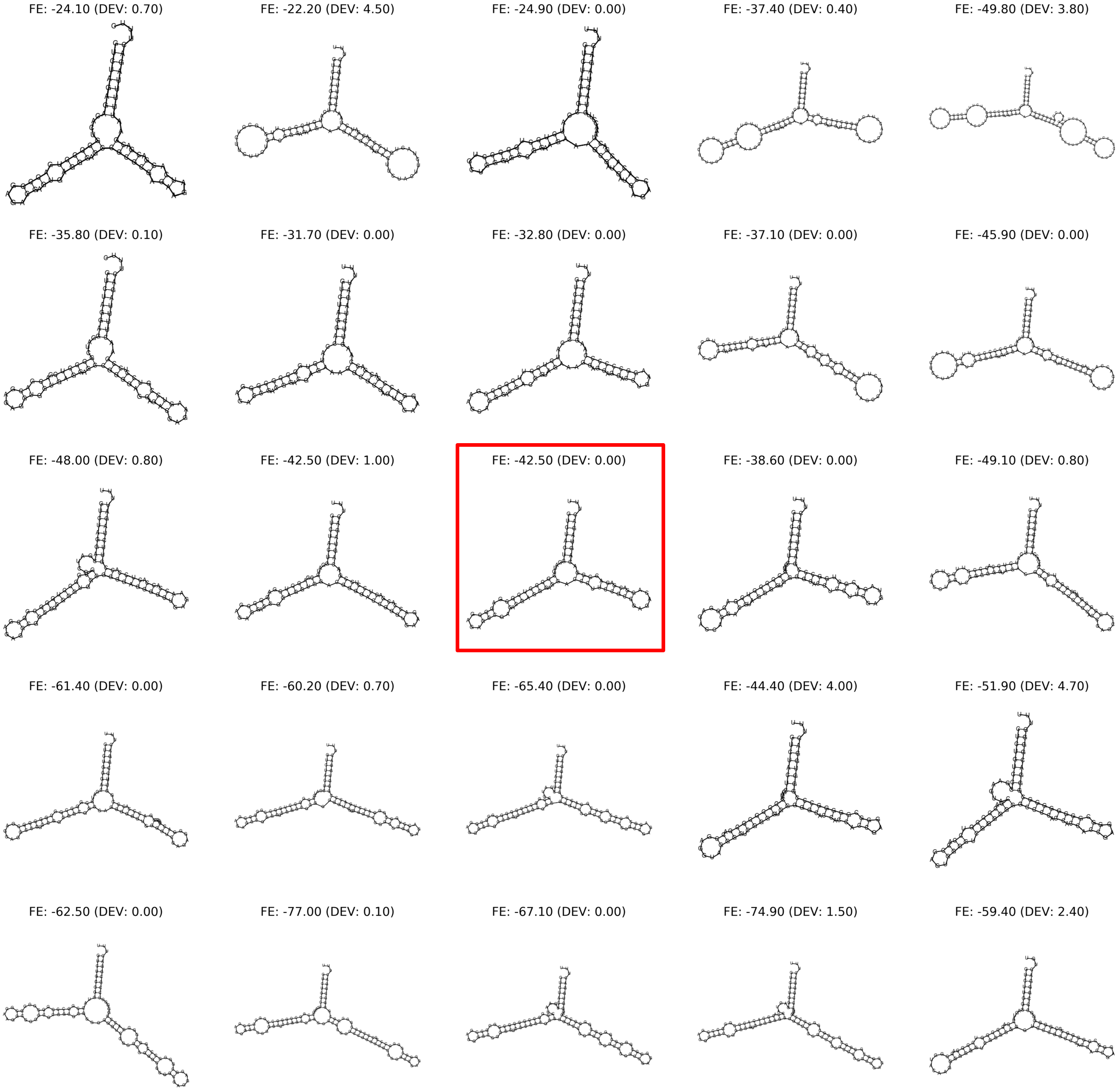}}
\caption{Neighborhood visualization of a 5S ribosomal RNA\protect\footnotemark which is marked by the red bounding box in the center and the walk in the latent space takes place on two random orthogonal axes. Note that actual secondary structure of this 5S ribosomal RNA plotted in the figure is different compared to the one deposited online due to the prediction of RNAfold.}
\label{fig:ribosomal}
\end{figure}
\footnotetext{\href{https://rnacentral.org/rna/URS000075B93F/9606}{https://rnacentral.org/rna/URS000075B93F/9606}}

\newpage
\section{Targeted RNA generation — an example}

\begin{figure}[!h]
\centerline{\includegraphics[width=0.95\textwidth]{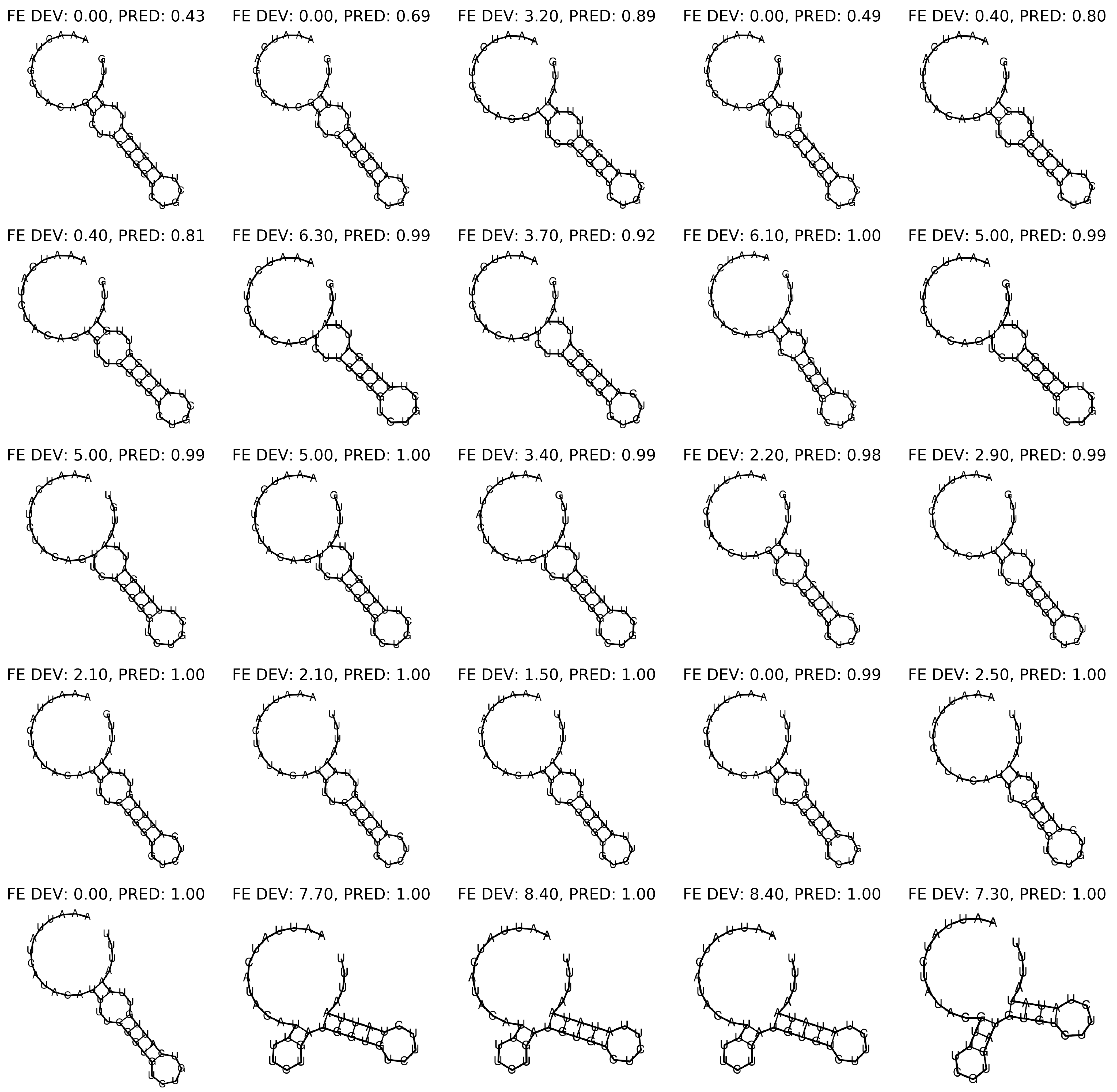}}
\caption{An example of searching novel structured RNAs with higher chance of binding to HuR. The optimization takes place in the latent space of HierVAE, starting from the initial encoding of a random RNA molecule in the test set, and at each step altering the latent encoding by using activation maximization on the embedding classifier. The trajectory of generated RNAs is shown in the order of left to right and top to bottom, and the field \textbf{PRED} indicates that the probability of binding, as predicted by another external full classifier on the decoded molecular structure, is overall increasing as the decoded RNA structures smoothly evolving.}
\label{fig:targeted-gen-example}
\end{figure}

\end{document}